\documentclass[aps,prl,twocolumn,showpacs,superscriptaddress,groupedaddress]{revtex4}  

\usepackage{graphicx}  
\usepackage{dcolumn}   
\usepackage{bm}        
\usepackage{amssymb}   
\usepackage{epsfig}
\usepackage{color}

\newcommand{\ea}{\textit{et al}}

\begin{document}

\title{High accuracy CO$_2$ line intensities determined from theory and
experiment}
\author{Oleg L. Polyansky},
\affiliation{Department of Physics and Astronomy, University College  London,
London WC1E 6BT, UK}
\affiliation{Institute of Applied Physics, Russian Academy of
Sciences, Ulyanov Street 46, Nizhny Novgorod, Russia 603950}

\author{Katarzyna Bielska}
\affiliation{Chemical Sciences Division, National Institute of Standards and
Technology, Gaithersburg, MD 20899, USA}
\affiliation{Institute of Physics, Faculty of Physics, Astronomy and Informatics,
Nicolaus Copernicus University in Torun, ul. Grudziadzka 5, 87-100 Torun,
Poland}

\author{M\'{e}lanie Ghysels}
\affiliation{Chemical Sciences Division, National Institute of Standards and
Technology, Gaithersburg, MD 20899, USA}

\author{Lorenzo Lodi}
\affiliation{Department of Physics and Astronomy, University College  London,
London WC1E 6BT, UK}

\author{Nikolai F. Zobov}
\affiliation{Institute of Applied Physics, Russian Academy of
Sciences, Ulyanov Street 46, Nizhny Novgorod, Russia 603950}

\author{Joseph T. Hodges}
\affiliation{Chemical Sciences Division, National Institute of Standards and
Technology, Gaithersburg, MD 20899, USA}

\author{Jonathan Tennyson}
\affiliation{Department of Physics and Astronomy, University College  London,
London WC1E 6BT, UK}
\email{j.tennyson@ucl.ac.uk.}

\date{\today}


\begin{abstract}

Atmospheric CO$_2$ concentrations are
being closely monitored by
remote sensing experiments which
rely  on knowing line
intensities with  an
uncertainty of 0.5\%\ or better.  Most available laboratory measurements have
uncertainties much larger than this.  We report
a joint experimental and theoretical study providing
rotation-vibration line intensities with the required accuracy.  The {\it ab
  initio} calculations are extendible to
all atmospherically important bands of CO$_2$ and to its isotologues.
As such they will form the basis for detailed CO$_2$
spectroscopic
line lists for future studies.
\end{abstract}

\maketitle

The quantity of carbon dioxide  (CO$_2$) in the earth's atmosphere and its role
in climate change has become a hotly debated topic both in scientific
and non-scientific circles. Several agencies are flying (for example GOSAT
\cite{GOSAT}, ACE \cite{ACE}, MIPAS \cite{MIPAS}, OCO-2 \cite{OCO}) or
preparing to launch (such as CarbonSat \cite{CarbonSat} and ASCENDS
\cite{ASCENDS}) experiments or even whole missions to explicitly monitor the
atmospheric CO$_2$
content or to do this as part of wider scientific programmes. Similarly,
international ground-based networks such as the total carbon column observation
network (TCCON) \cite{TCCON} and the Network for the Detection of Atmospheric
Composition Change (NDACC) \cite{NDACC} are also dedicated to monitoring
atmospheric CO$_2$. A major aim of this activity is to establish CO$_2$
concentrations at the parts per million (ppm) level or, preferably, better.
These projects aim not only to look at global CO$_2$ levels
and their variations, but also at sources and sinks of CO$_2$. This activity is
clearly vital to monitoring and hopefully controlling the anthropic greenhouse
effect due to CO$_2$ and hence climate change.

All CO$_2$ remote sensing activities, from both the ground and space,
rely on monitoring CO$_2$ vibration-rotation spectra. They are
therefore heavily dependent on laboratory spectroscopy for reliable
parameters which are essential for interpreting the atmospheric
spectroscopic data.  These parameters are of three types: line
centres, line profiles and line intensities. Line centres or positions
are established to high accuracy in many laboratory high resolution
spectroscopy studies and in general do not require significant
improvement for studies of the earth's atmosphere. Line profiles are
more difficult to determine, but significant progress on these has
been made in recent years with, for example, the inclusion of line
mixing in both the HITRAN database \cite{jt557} and many retrieval
models.  Here we specifically focus on the issue of determining
accurate line intensities. We present first results from a newly-developed
experiment designed to measure line intensities with an uncertainty of 0.3~\%,
and a new {\it ab initio} model also designed to achieve this level of accuracy.

Accurate line intensities are crucial for a successful retrieval since they
relate directly to the CO$_2$ column being retrieved.  Without high accuracy
values for line
intensities, reliable retrievals are simply not possible. If current missions are to
fulfil their goals intensities accurate to 0.3 -- 0.5~\%\ are really required \cite{XOCO}.
The laboratory procedures used up to now 
simply do not give this level of accuracy and current retrievals values are
limited by the available laboratory data
\cite{08CoBoTo.CO2,11WuViJo.CO2,14SiBoNa.CO2}.
Data are required not only for the main isotopologue, $^{12}$C$^{16}$O$_2$, but
also for isotopically substituted species such as $^{14}$C$^{16}$O$_2$, which
can be used to monitor recently added fossil-fuel-derived carbon emissions
\cite{13LeMiWo.CO2}.

It is much harder to accurately measure line intensities than line positions in
the laboratory. Typical uncertainties for experimental
line intensity data used in atmospheric models and retrievals are 3 to 10~\%\
\cite{04WaPeTa.CO2,08PeCaGa.CO2,10SoKaTa.CO2} and even high quality measurements
(eg Boudjaadar \ea \cite{06BoMaDa.CO2}) usually only provide accuracies in the 1
to 3~\%\ range, still significantly
worse than required for precision remote sensing. There are three published studies aimed at
measuring CO$_2$ line intensities with an accuracy
better than 1~\%\ \cite{07CaPaCa.CO2,09CaWeCa.CO2,11WuViJo.CO2}. However these
studies only considered a small set of lines, in the case of Wuebbler
{\it et al} \cite{11WuViJo.CO2} only a single line, and do not agree
with each other within their given uncertainties.  On the theory side Huang {\it
et al} have recently performed a comprehensive treatment of
the CO$_2$ vibration-rotation spectrum
\cite{12HuScTa.CO2,13HuFrTa.CO2,14HuGaFr.CO2} using theoretical procedures
similar to those employed here. These studies produced an
excellent spectroscopically determined potential energy surface (PES), which we
use below, but had a more limited goal for the accuracy of their intensities of
between 3 and 5~\%.

The aim of this work is to provide an accurate theoretical solution to the
problem of CO$_2$ line intensities based on the application of
high-accuracy, {\it ab initio} quantum mechanical calculations tested against 
laboratory measurements of unprecedented low uncertainty. An advantage of our calculations is that they
can be applied not only to all bands of importance but also to all isotopically
substituted variants of the molecule with essentially uniform precision. The
disadvantage of {\it ab initio} methods has been traditionally that they are
hard to perform to high accuracy and it is
difficult to estimate their uncertainty.  Here we present a joint experimental
and theoretical study demonstrating an {\it ab initio} theoretical model capable
of reproducing line intensities of $^{12}$C$^{16}$O$_2$ with a combined uncertainty of
about 0.3\%.

Theoretically, the intensity of a given spectral line is directly proportional
to the
square of the transition dipole, \(\mu_{if} = \int \Psi_i \underline{\mu} \Psi_f
\mathrm{d}\tau\), where
$\Psi_i$ and $\Psi_f$ are the initial and final wavefunction. For a
vibration-rotation
transition, $\underline{\mu}$ is the dipole moment surface (DMS) and the
integration runs over all coordinates of the nuclei.
The accurate calculation of molecular line intensities requires an accurate DMS
and an accurate potential energy surface (PES) to provide
wavefunctions. The rotation-vibration Schr{\"o}dinger equation is solved numerically
to compute wavefunctions; we use the DVR3D program suite \cite{jt338} for that
purpose. Experience shows that best results are obtained by combining a
spectroscopically determined PES with a fully {\it ab initio} DMS \cite{jt509}.

Here we use the highly accurate, empirical, CO$_2$ PES of Huang {\it et
  al} \cite{12HuScTa.CO2}. What we require is an extra-high-accuracy
CO$_2$ DMS. Systematic studies of the DMS of water \cite{jt424,jt509} have
shown that sub-1~\%\ accuracy requires consideration of many
effects neglected in standard {\it ab initio} treatments and that
there is a strong correlation between the accuracy of the DMS and the
underlying PES associated with it \cite{jt509}.  A number of studies
\cite{12HuScTa.CO2,jt512} have also demonstrated the importance of
generating the DMS using a dense grid of points. As shown for water
\cite{jt309,jt550}, the key ingredients for a high-accuracy {\it ab initio}
treatment involve the use of multireference configuration interaction (MRCI)
calculations with large basis
sets (5 or 6-zeta quality) and of large active spaces.
It is furthermore necessary to add various corrections
due to relativity (and even quantum electrodynamics)
and failure of the Born-Oppenheimer (BO) approximation.

Water is a 10 electron system which lends itself to large systematic
calculations. These calculations scale combinatorially and therefore are not 
currently computationally feasible for the
22-electron CO$_2$ molecule. For this reason it was necessary
to design a new model: preliminary test calculations with this model,
detailed in the supplementary material, were performed for
CO.

Our calculations used the MOLPRO \cite{12WeKnKn.methods} quantum
chemistry package to calculate the PES and DMS of CO$_2$ at about 2000
randomly-selected points with energies up to about 15~000 cm$^{-1}$
above the minimum. The calculations used all-electron
MRCI and the aug-cc-pwCVQZ basis set. Relativistic corrections
were determined  from the one-electron mass-velocity-Darwin (MVD1) term
and fitted separately. The DMS was determined using finite-field effects
rather than as an expectation value. As CO$_2$ is heavier
and more rigid than water, it transpired that non-BO corrections
are of lesser importance, as has been shown before \cite{12HuScTa.CO2},
and were not included. Full details of the calculation, including
analytic representations of our final {\it ab initio} DMS, which were obtained
as polynomial expansions in symmetry coordinates,
and the associated PES are given in the
supplementary material.

DVR3D  calculations for the three fundamental bands of
$^{12}$C$^{16}$O$_2$ using our \textit{ab initio} PES show that
the discrepancy between calculated and observed
energy levels is about 1 cm$^{-1}$, almost an order of
magnitude smaller than the best previous \textit{ab initio} calculations
\cite{12HuScTa.CO2}.  This level of accuracy for the energy levels
should be a pointer towards the accuracy of the corresponding DMS  and suggests
that the intensity of
strong and medium lines should be predicted to within 0.5~\%.

We made the most accurate measurements ever reported for CO$_2$ line intensities using the 
frequency-stabilized cavity ring-down spectroscopy (FS-CRDS) \cite{hodges04,hodges05,long12} technique.  
FS-CRDS is a high-accuracy method that yields absorption spectra in terms of known integer multiples of
 the longitudinal mode spacing of a resonant optical cavity ($x$-axis) and observed cavity decay times
 ($y$-axis). In contrast to other absorption spectroscopy methods, this approach is immune to intensity 
and frequency fluctuations in the probe laser, consists of a relatively compact sample volume and does 
not require explicit determination of the absorption path length.  Our gas samples consisted of a known
 molar fraction of CO$_2$ in air with values that were referenced to gravimetrically prepared primary 
standard mixtures.  Further, to mitigate exchange of CO$_2$ with internal surfaces of the spectrometer, 
the sample gas was continuously introduced into the absorption spectrometer.  
Using this approach we measured the spectroscopic areas of 27 vibration-rotation transitions of CO$_2$ 
in the wave number region 6200 - 6258 cm$^{-1}$.  These spectroscopic features correspond 
to $P$- and $R$-branch transitions of the $^{12}$C$^{16}$O$_2$ (30013)-(00001) vibrational band.  

The individual transitions were probed using the frequency-agile, rapid scanning spectroscopy (FARS) method \cite{truong13} to rapidly and precisely tune the laser in a stepwise fashion through successive cavity resonances.  To this end, we used a high-finesse ($\sim$1.6$\times$10$^{5}$), 75-cm-long cavity ring-down spectrometer whose length was actively stabilized with respect to a frequency-stabilized HeNe laser having a drift (on the time scale of spectrum acquisition) of less than 0.5 MHz.  Two continuous-wave, distributed-feedback diode (DFB) lasers (1 MHz nominal line width) provided the wavelength coverage required to interrogate all the CO$_2$ transitions reported.

Relative standard deviations in the absorption coefficient, $\alpha_{\rm tot}$, at a given frequency, $\nu$, were 0.08 $\%$, and over the entire 10 GHz spectral
window spectra were acquired in $\sim$45 s giving spectrum signal-to-noise (SNR) ratios of $\sim$5000:1.  Between 20 and 100 spectra were acquired at pressures corresponding to 6.7 kPa, 13.3 kPa and 20 kPa. Typically, we fit each observed line with the sum of a linear baseline and multiple quadratic speed-dependent Nelkin-Ghatak profiles (qSDNGPs) \cite{ciurylo98} (one for each observed line) to the etalon-subtracted spectrum ($\alpha_{\rm tot}(\nu)$). This analysis gave the fitted area $A$($p$, $T$) at each pressure, $p$, and temperature, $T$. Here we report the average, measured, pressure-independent line intensity $S$.

Gas pressure and cell temperature were actively stabilized to minimize drift during the acquisition of each spectrum, and the measurements of $p$ and $T$ were traceable to NIST primary standards.  Nominal relative standard uncertainties (RSUs) considered include: spectrum tuning step size (0.001 $\%$), statistical uncertainty in fitted area (0.01 $\%$), pressure (0.015 $\%$), systematic residual area uncertainty (0.02 $\%$), uncertainty in isotopic composition (0.02 $\%$), gas temperature (0.05 $\%$), variations in baseline etalons (0.05-0.9
$\%$) and sample molar fraction (0.07 $\%$).  Adding these components in quadrature gives RSU values ranging from 0.1 $\%$ - 1$\%$, with the strongest lines being relatively insensitive to baseline uncertainties and having RSUs between 0.1 $\%$ and 0.2 $\%$.  We attribute these exceptionally low combined uncertainties to the high fidelity and sensitivity of the FS-CRDS method and to the accurately known CO$_2$ concentration in the sample gas.

Figure 1 compares literature values for the (30013) -- (00001) band  of $^{12}$C$^{16}$O$_2$ with the present {\it ab initio} calculations and our  measurements. These results are summarized as the relative
intensity difference versus rotational quantum number $m$.  The literature results include two independent sets of measurements reported by Boudjaadar et al. \cite{06BoMaDa.CO2} (LPPM and GSMA) and other intensities given in HITRAN 2012  \cite{jt557}.  These three data sets are based on Fourier transform spectroscopy measurements of pure carbon dioxide samples. We also include intensities archived in the carbon dioxide spectral databank (CDSD) \cite{11TaPe.CO2}, which are based on fits to several spectroscopic measurements. When taken together, the relative intensity differences have a standard deviation and mean of 0.9 $\%$, and 0.03~$\%$, respectively. We find the smallest root-mean-square deviation (0.23 $\%$) and minimum absolute relative difference (0.33 $\%$) when comparing the present calculations to our measurements. These results confirm that the relative uncertainties of the present {\it ab initio} intensity calculations and measurements are in good agreement and well below the percent level, which constitutes a substantial improvement over previous intensity measurements.

\begin{figure}
\includegraphics[angle=0,width=10 cm,clip]{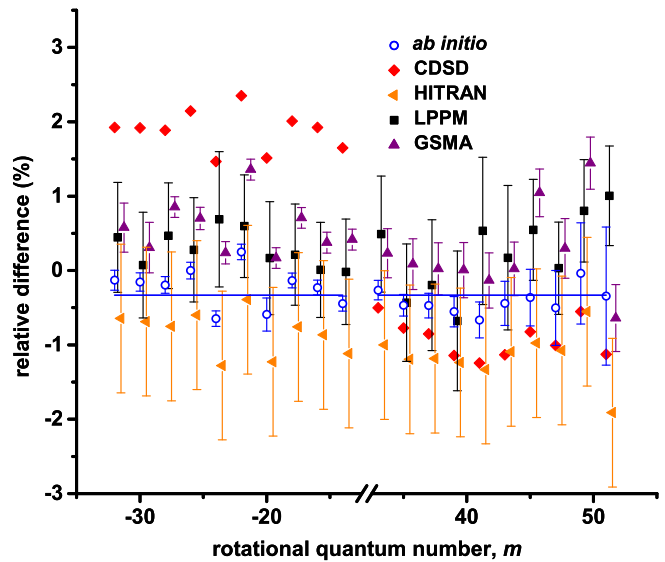}
\caption{\protect{{\it Ab initio} intensities (this work) and literature values
relative to the present measurements versus rotational quantum number $m$.
Error bars represent $\pm$ 1 standard deviation. The average, weighted (based on
the inverse of the RSU$^2$) relative difference between the calculations and the
present measured values equals -0.33 $\%$ (indicated by the horizontal line) and
has a standard error of 0.05 $\%$.  This comparison gives a root-mean-square
deviation of 0.23 $\%$ consistent with the measurement RSU. Data sources
are CDSD \cite{11TaPe.CO2},  HITRAN  \cite{jt557}, LPPM \cite{06BoMaDa.CO2} and
GSMA \cite{06BoMaDa.CO2}.}}
 \label{Fig1}
\end{figure}

Intensities of the  (20012) -- (00001) band of $^{12}$C$^{16}$O$_2$ has also
been the subject of precision studies.
As shown in Fig.~2, only one of the intensities measured by  Casa {\it et al}
\cite{07CaPaCa.CO2,09CaWeCa.CO2} is within 0.3 \%\
of our calculations, while the average relative difference is greater than 1 \%.
However, for one of these problematic lines, R(12),
the intensity has been remeasured independently by Wuebbeler {\it et al}
\cite{11WuViJo.CO2}.
This remeasured value is within 0.2~\%\ of our calculated value.
It would appear that  the intensities of Casa {\it et al} are measured less
accurately than claimed.

\begin{figure}
\includegraphics[angle=0,width=10 cm,clip]{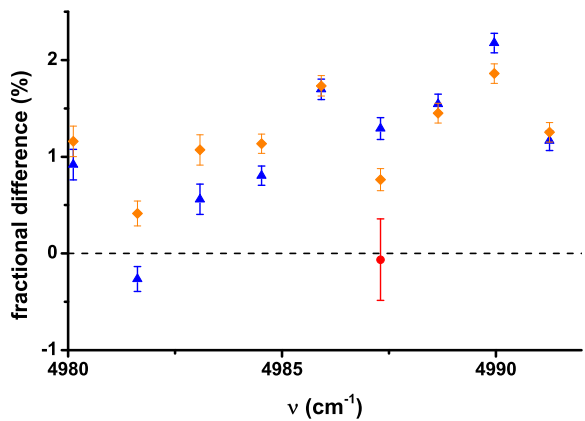}
\caption{ {\it Ab initio} line intensities (this work) for the  (20012) --
(00001)  band of the $^{12}$C$^{16}$O$_2$ molecule at 296~K
compared to the measurements of Casa \ea\ in 2007  \cite{07CaPaCa.CO2} (blue
triangles) and 2009 \cite{09CaWeCa.CO2} (orange
diamonds), and
the single 2011 measurement of Wuebbeler \ea\ \cite{11WuViJo.CO2} (red circle).}

 \label{Fig2}
\end{figure}

Table 1 compares our calculated intensities with those given in HITRAN for
different R(10) transitions: these results
are typical of comparisons with other transitions.   Good agreement, within the
rather
large HITRAN uncertainties, is found in all cases. In particular,
we note that very good agreement, to about 0.4~\%\ and 0.2~\%\ respectively, is
obtained for transitions within the two fundamental bands,
(01101) -- (00001) and (00011) -- (00001) (fuller results are given in the
supplementary material).
We suggest that this is not a coincidence. These HITRAN intensities are the
result of calculations using a fitted, effective DMS \cite{99TaPeTe.CO2}.
As the constants of the effective DMS responsible for these strong fundamental
bands are determined by fits to both these and
over 100 other bands, the large amount of data available from a wide variety of
experiments leads to an overall accuracy of the line
intensities which should significantly improve on the average 2-5~\%\
uncertainty in the experimental data used in the fit.
The excellent agreement demonstrates both the extreme accuracy of our
calculations and the effectiveness of the Hamiltonian fit in this case.
Furthermore, for all bands, the relative differences between the HITRAN and
calculated intensities are generally much smaller
in magnitude than the rather large HITRAN uncertainties given in Table 1.  We
suggest that our new intensities generally represent a
considerable improvement in accuracy and
that the HITRAN uncertainties may be reduced substantially by taking into
account the present calculations.

\begin{table}
\caption{Line intensities, in cm/molecule, of the R(10) transitions of the main
bands of  $^{12}$C$^{16}$O$_2$ at 296 K. Our calculations (C) versus HITRAN (H)
\cite{jt557}; ``Unc" gives the stated HITRAN uncertainty for each line.
(Powers of 10 are in parentheses)}
\begin{tabular}{c r  c crr}
\hline\hline
Band&$\tilde{\nu}$/cm$^{-1}$& $I$(H)& $I$(C)&(H-C)/H (\%) &
Unc.(\%) \\
\hline
(01101) &  676.01  & 1.524($-19$) & 1.519($-19$) &  0.35 &    5  \\
(11102) & 1941.12  & 7.262($-25$) & 5.975($-25$) & 17.73 &   20 \\
(11101) & 2085.46  & 2.576($-23$) & 2.571($-23$) &  0.19 &   10 \\
(00011) & 2357.32  & 3.116($-18$) & 3.117($-18$) & -0.04 &    5  \\
(21103) & 3190.15  & 1.302($-25$) & 1.284($-25$) &  1.42 &   20 \\
(21102) & 3347.91  & 1.694($-24$) & 1.721($-24$) & -1.61 &   20 \\
(21101) & 3509.27  & 1.071($-24$) & 1.108($-24$) & -3.47 &   20 \\
(10012) & 3621.06  & 3.351($-20$) & 3.395($-20$) & -1.31 &    5  \\
(02211) & 3667.64  & 1.765($-25$) & 1.787($-25$) & -1.23 &   20 \\
(10011) & 3722.94  & 5.162($-20$) & 5.228($-20$) & -1.27 &    5  \\
(31104) & 4424.87  & 3.062($-27$) & 3.356($-27$) & -9.61 &   20 \\
(31103) & 4599.65  & 3.865($-26$) & 3.986($-26$) & -3.14 &   20 \\
(31102) & 4761.96  & 5.544($-26$) & 5.687($-26$) & -2.58 &   20 \\
(20013) & 4861.93  & 2.370($-22$) & 2.370($-22$) &  0.00 &    5  \\
(12212) & 4896.40  & 4.070($-27$) & 3.972($-27$) &  2.40 &   20 \\
(31101) & 4946.98  & 3.061($-27$) & 2.362($-27$) & 22.85 &   20 \\
(20012) & 4985.93  & 1.127($-21$) & 1.156($-21$) & -2.60 &    5  \\
(12211) & 5070.13  & 4.949($-27$) & 4.889($-27$) &  1.21 &   20 \\
(30013) & 6236.03  & 1.515($-23$) & 1.551($-23$) & -2.38 &   10 \\
\hline
\end{tabular}
\end{table}

We present new high accuracy measurements and {\it ab initio} calculations of transition intensities for the key $^{12}$C$^{16}$O$_2$ molecule. Agreement between calculations and experiments is at the 0.3 $\%$ level, which represents a significant improvement over previous theoretical and measured values.  Our theoretical procedure is capable of producing comprehensive line lists not only for $^{12}$C$^{16}$O$_2$ but also for the various isotopically substituted versions. These line lists, for which experimental line frequencies can be used, will give a significant and important improvement in CO$_2$ line intensities available for atmospheric remote sensing and other studies. These line lists will be presented elsewhere.

\section*{Acknowledgement}
This work was supported by UK Natural Environment Research Council under grant
NE/J010316, ERC Advanced Investigator Project 267219, the Russian Fund for
Fundamental Studies, and the Climate Sciences Measurements Program of the
National Institute of Standards and Technology and state project IAP RAS
No.0035-2014-009.


\end{document}



\baselineskip24pt

\maketitle

\section*{Supplementary information}

\subsection*{Calculations}

The overall methodology used for these calculations has been described
in detail in a tutorial review by two of us \cite{jt475}.

{\it Ab initio} calculations on
isotopologues of water \cite{jt550} suggest that 9 factors need to be
taken into consideration if results approaching spectroscopic accuracy
are to be achieved.  These components, in approximate decreasing
order of importance, are: 1) The treatment of
electron correlation at the multi reference configuration interaction
(MRCI) level of theory; 2) Use of the highest possible one-particle
basis set, normally at the aug-cc-pCV6Z level or [5,6]z complete basis
set (CBS) extrapolation level; 3) use of a dense grid of points; 4)
the complete active space (CAS) used in the MRCI calculations should
be chosen to be  larger than the MOLPRO default which is generally the full valence one; 
5) Core-valence correlation taken into account explicitly; 
6) Relativistic
corrections  computed at an appropriate
level must be used; 7) Born-Oppenheimer diagonal
correction (BODC); 
8) Quantum electrodynamics corrections, which have been
shown to change vibrational levels of water molecule by more than 1
cm$^{-1}$ \cite{jt265}, should be considered. 9) Nonadiabatic
correction have to be considered.

We have tested all these 9 factors recently when performing \ai\
predictions of the energy levels and transition intensities of the CO
molecule \cite{coinprog}.  An energy level accuracy of 0.1 cm$^{-1}$
or better was achieved for vibration-rotation levels up to 30 000
cm$^{-1}$ above vibration-rotation ground state.  
As CO has only 14 electrons, a high level of \ai\ theory,
as yet inaccesible for CO$_2$, could be employed. We then tried many
different combinations of the above 9 factors with smaller basis sets
in order to evaluate the accuracy of such reduced calculations.  We
found that relatively-simple purely-BO calculations at the MRCI
all-electron level with an aug-cc-pwCV5Z basis set give very accurate
results (better then 1 cm$^{-1}$) for the ro-vibrational energy
levels. The dipole moment surface (DMS) calculated at this level of theory
resulted in calculated intensities for strong and
medium transitions which are  very close to those given by the
calculation which converged in full all 9 factors.

As CO is in many respects similar to CO$_2$, we have tried to transfer this
analysis to the CO$_2$ molecule. We find that the CO$_2$ MRCI, all-electron,
BO PES obtained with aug-cc-pwCVQZ basis set resulted in
energy levels within 1 cm$^{-1}$ of the fundamental band origins for CO$_2$.
Experience shows that this level of agreement for the PES leads to dipole transition
intensities which are within 0.5 \%\ of the observed value for medium and strong lines.
We decided to
employ this level of theory for the calculation of the CO$_2$ DMS described in the
main body of this paper. We ascribe an uncertainty of
0.5 \%\ to the intensities arising from these calculations.

Electronic structure calculations were performed
in the range $1.1 < (r_1,r_2) < 1.45$~\AA, and   $135^\circ < \theta < 180^\circ$,
where $(r_1,r_2,\theta)$ are the standard bond-length, bond-angle coordinates. Only
points lying below 15~000 cm$^{-1}$ above the bottom of the well
were used to determine the DMS.
MRCI calculations were performed using the aug-cc-pwCVQZ basis. The six core 1s electrons were frozen
and the remaining 16 were distributed amongst the 12 (2s,2p) orbitals in the
active space.
Relativistic calculations corrections were included pertubatively at the
one-electron mass-velocity Darwin (MVD1) level.
The dipoles were computed as the derivative
of the energy, considered as a function of an external, weak,
uniform electric field, for zero field strength.
For each geometry a total of 4 independent MRCI
runs were performed in the presence of electric field of an 0.0003 a.u.
in the positive and negative x and y directions to
obtain finite-difference dipoles.

Several surfaces were fitted (each is provided separately as a Fortran
program).  The main, non-relativistic PES was fitted using 50
constants which reproduced the points with a standard deviation,
$\sigma =1.54$ cm$^{-1}$.  A relativistic correction surface was
produced separately in which 31 constants reproduced 1295 points with
$\sigma =0.563$ cm$^{-1}$.

The linear CO$_2$ molecule was taken to lie along the y axis. Fits were performed separately
for the parallel (y) and perpendicular (x) DMS using the coordinates:
\begin{equation}
 S_1 = (r_1+r_2)/2-r_e,\ \
  S_2 = (r_1-r_2)/2-r_e,\ \
S_3 = 180 - \theta
\end{equation}
where $r_e$ is the equilibrium C--O bond length of 1.16 \AA.
Each DMS component was expanded as a power series of the form $S_1^\ell S_2^m S_3^n$.
By symmetry the x component has $m$ even and $n$ odd only, and the y component has
$m$ odd and $n$ even. All $\ell \geq 0$ are allowed in both cases. Fits include terms up to fifth-order
and considered energy up 15~000 cm$^{-1}$ above the minimum and excluded points were the dipole
component was zero by symmetry. For x, 1963 points were fitted by 17 constants
with $\sigma =2.25 \times 10^{-5}$ a.u.;
for y, 1433 points were fitted by 19 constants with $\sigma =1.85 \times 10^{-5}$ a.u.

DVR3D calculations for $^{12}$C$^{16}$O$_2$ were performed using nuclear masses and in
symmetrized Radau coordinates with the so-called bisector embedding,
which places the z-axis along the molecule at equilibrium. 120 points
based on Gauss-(associated) Legendre quadrature were used for the
angular basis. 100 Gauss-Laguerre grid points based on spherical
oscillator functions \cite{jt23} with $\alpha = 0$ and $\omega_e =
0.005$ a.u. were used for each radial coordinate. Solutions were
obtained by diagonalising vibrational Hamiltonians of maximum
dimension 9500 and vibration-rotation Hamiltonians of size $(J+1) \times 600$.


\subsection*{Experiment}

Gas samples
comprised synthetic, Ar-free, air mixtures containing 427.2 nmol
mol$^{-1}$ of carbon dioxide with a relative standard uncertainty
(RSU) in concentration of 0.07 $\%$.  Prior to the CRDS analysis, the
sample gas composition was referenced via gas chromatography to four
gravimetrically prepared, primary standard mixtures of total
carbon-dioxide-in-air, each of which had a RSU of $\sim$0.015 $\%$.
Because of the petrochemical source of the carbon dioxide used in the
synthetic air mixture, the relative isotopic abundance of
$^{12}$C$^{16}$O$_2$ was assumed to be $I_{\rm 626}$ = 0.98460,
corresponding to $\delta$$^{13}$C$_{\rm VPDB}$ = -4.16 $\%$ and
$\delta$$^{18}$O$_{\rm VPDB}$ = -2.37 $\%$ \cite{verkouteren99}.
To minimize adsorption/desorption of the CO$_2$ with
the internal surfaces of the ring-down cell, spectra were acquired
with a steady flow (30 cm$^3$ min$^{-1}$) of sample gas near room
temperature.

In the frequency-agile, rapid
scanning spectroscopy (FARS) method \cite{truong13}, tunable sidebands
are impressed upon the probe laser beam which propagates through a
microwave-driven, waveguide electro-optic phase modulator (EOM).  This
approach exploits the frequency selectivity of the ring-down cavity to
block the laser carrier frequency and all but one of the microwave
sidebands so that specific cavity resonances can be selectively and
rapidly probed.  FARS results in a linear and accurate spectrum
detuning axis with an average value that depends on the
Cs-clock-referenced, microwave driving frequency and the stability of
the ring-down cavity length. To augment laser throughput, we used a
booster-optical amplifier (BOA), which yielded 10 mW of incident
optical power in the first-order sideband.  The FARS-BOA scheme
overcomes the relatively slow thermal tuning of the DFB laser, enables
sequential mode-by-mode interrogation of the ring-down cavity, and
provides high beam extinction ratios \cite{long13} necessary to
optimize CRDS measurement statistics. At each laser detuning $\nu$, 50
ring-down signals were measured in transmission ($\sim$5 $\mu$W peak
power) with a DC-700 kHz bandwidth InGaAs detector, then digitized at
16-bits of resolution and fit in real-time with an exponential
function to give the spectrum of total cavity losses $\alpha_{\rm
  tot}(\nu)$, (equal to the cavity baseline plus absorption
coefficient).

 In addition to the target transition, most spectra also
contained weak, interfering lines from various isotopologues of carbon
dioxide as well as etalon features caused by coupled-cavity effects
\cite{courtois13}. We characterized these etalons by measuring the
empty-cavity spectrum.

All measured results are divided by the relative isotopic abundance
(0.98460) and normalized to the reference temperature $T_r$=296 K by
$A$($T$)[$k_B$$T$/($pI_{\rm 626}$)][$S$($T_r$)/$S$($T$)], where $k_B$
is the Boltzmann constant and $S$($T_r$)/$S$($T$) depends on the lower
state energy and partition function.
All FS-CRDS measurements reported here were made at the National Institute of Standards and Technology 
(NIST) in Gaithersburg, Maryland.

\subsection*{Results}

The comparison with HITRAN given in Table 1 in the main paper is for a set of R(10) lines
since, at 296 K, the intensity of these lines shows particularly weak sensitivity to temperature
effects. Tests performed for other lines within the bands shown give very similar results. 

Numerical values for our results are given in Tables 1 -- 3. Note results for table 1
are given for pure $^{12}$C$^{16}$O$_2$, which means any factors due to isotopic abundances have been removed, whereas
other tabulated values are for $^{12}$C$^{16}$O$_2$ in natural abundance.

\begin{table}
\caption{Line intensities$^a$, in cm/molecule, of $^{12}$C$^{16}$O$_2$  obtained experimentally and
theoretically in this work for the (30013) -- (00001) band at 296~K
The "Obs" values given below were obtained by dividing the measured intensities by the assumed 
isotopic abundance of 0.984599.}
\begin{tabular}{c l c c r}
\hline\hline
$\tilde{\nu}$/cm$^{-1}$  & Transition& $I$(Obs)& $I$(Calc)&(Obs-Calc.)/Obs. (\%) \\
\hline
6199.44 & P(32) & 7.5205E-24 & 7.5107E-24 & 0.13  \\
6201.44 & P(30) & 8.9295E-24 & 8.9158E-24 & 0.15  \\
6203.40 & P(28) & 1.0401E-23 & 1.0381E-23 & 0.20  \\
6205.34 & P(26) & 1.1847E-23 & 1.1847E-23 & 0.00  \\
6207.25 & P(24) & 1.3328E-23 & 1.3242E-23 & 0.65  \\
6209.12 & P(22) & 1.4444E-23 & 1.4480E-23 & -0.25 \\
6210.97 & P(20) & 1.5564E-23 & 1.5472E-23 & 0.59  \\
6212.79 & P(18) & 1.6146E-23 & 1.6124E-23 & 0.13  \\
6214.59 & P(16) & 1.6389E-23 & 1.6351E-23 & 0.23  \\
6216.35 & P(14) & 1.6153E-23 & 1.6081E-23 & 0.45  \\
6218.09 & P(12) & 1.5319E-23 & 1.5263E-23 & 0.36  \\
6219.80 & P(10) & 1.3963E-23 & 1.3877E-23 & 0.62  \\
6221.48 & P(8)  & 1.1997E-23 & 1.1933E-23 & 0.54  \\
6223.13 & P(6)  & 9.5325E-24 & 9.4774E-24 & 0.58  \\
6224.75 & P(4)  & 6.6246E-24 & 6.5921E-24 & 0.49  \\
6247.46 & R(28) & 1.1244E-23 & 1.1216E-23 & 0.25  \\
6248.58 & R(30) & 9.7011E-24 & 9.6393E-24 & 0.64  \\
6249.67 & R(32) & 8.1491E-24 & 8.1276E-24 & 0.26  \\
6250.73 & R(34) & 6.7582E-24 & 6.7267E-24 & 0.47  \\
6251.76 & R(36) & 5.4928E-24 & 5.4669E-24 & 0.47  \\
6252.76 & R(38) & 4.3888E-24 & 4.3645E-24 & 0.55  \\
6253.73 & R(40) & 3.4466E-24 & 3.4237E-24 & 0.66  \\
6254.67 & R(42) & 2.6514E-24 & 2.6397E-24 & 0.44  \\
6255.57 & R(44) & 2.0080E-24 & 2.0007E-24 & 0.37  \\
6256.45 & R(46) & 1.4985E-24 & 1.4910E-24 & 0.50  \\
6257.29 & R(48) & 1.0932E-24 & 1.0928E-24 & 0.04  \\
6258.10 & R(50) & 7.9025E-25 & 7.8753E-25 & 0.34  \\
\hline
\end{tabular}
\end{table}

\begin{table}
\caption{Line intensities, in cm/molecule,  of the   (20012) -- (00001) band of         
 $^{12}$C$^{16}$O$_2$ molecule at 296~K. Observed results are
due to Casa {\it et al} in 2007 \cite{07CaPaCa.CO2} and in 2009
\cite{09CaWeCa.CO2}, except for line marked $^*$ which was measured in 2011
by Wuebbeler {\it et al} \cite{11WuViJo.CO2}. Differences, D(year), are (Obs -- Calc.)/Obs for
given measurement.  }
\begin{tabular}{c l c c cc cc}
\hline\hline
$\tilde{\nu}$/cm$^{-1}$  & Transition& $I$(Obs2007)&  $I$(Obs2009)&$I$(Calc)& D(2007) (\%) & D(2009) (\%)\\
\hline
4980.12 & R(2)  & 3.7890E-22 & 3.7800E-22 & 3.8238E-22 & -0.88 & -1.12 \\
4981.62 & R(4)  & 6.2290E-22 & 6.1870E-22 & 6.2126E-22 &  0.30 & -0.37 \\
4983.08 & R(6)  & 8.3050E-22 & 8.2630E-22 & 8.3515E-22 & -0.52 & -1.03 \\
4984.52 & R(8)  & 1.0074E-21 & 1.0041E-21 & 1.0155E-21 & -0.77 & -1.10 \\
4985.92 & R(10) & 1.1370E-21 & 1.1366E-21 & 1.1563E-21 & -1.65 & -1.69 \\
4987.30 & R(12) & 1.2379E-21 & 1.2444E-21 & 1.2539E-21 & -1.25 & -0.72 \\
4987.30 & R(12)$^*$& & 1.2547E-21 & 1.2539E-21 &   &  0.11 \\                                    
4988.65 & R(14) & 1.2877E-21 & 1.2889E-21 & 1.3076E-21 & -1.50 & -1.41 \\
4989.96 & R(16) & 1.2912E-21 & 1.2952E-21 & 1.3193E-21 & -2.13 & -1.82 \\
4991.25 & R(18) & 1.2780E-21 & 1.2769E-21 & 1.2929E-21 & -1.12 & -1.21 \\
\hline
\end{tabular}
\end{table}

\begin{table}
\caption{{\it Ab initio} line intensities (this work) of the (01101) -- (00001) band (00011) -- (00001) bands
of $^{12}$C$^{16}$O$_2$ at 296~K
compared to HITRAN \cite{jt557}.}
\begin{tabular}{c l c c cc}
\hline\hline
$\tilde{\nu}$/cm$^{-1}$  & Transition& $I$(HITRAN)& $I$(Calc)& (HITRAN-Calc/HITRAN) (\%) \\
\hline
 648.88 & P(24) & 1.028E-19 & 1.024E-19 & 0.41  \\
 650.40 & P(22) & 1.130E-19 & 1.125E-19 & 0.46  \\
 651.93 & P(20) & 1.211E-19 & 1.206E-19 & 0.40  \\
 653.46 & P(18) & 1.265E-19 & 1.260E-19 & 0.41  \\
 654.99 & P(16) & 1.283E-19 & 1.278E-19 & 0.39  \\
 656.53 & P(14) & 1.259E-19 & 1.254E-19 & 0.38  \\
 658.06 & P(12) & 1.188E-19 & 1.184E-19 & 0.37  \\
 659.61 & P(10) & 1.067E-19 & 1.064E-19 & 0.31  \\
 661.15 & P(8)  & 8.981E-20 & 8.951E-20 & 0.33  \\
 662.70 & P(6)  & 6.835E-20 & 6.813E-20 & 0.32  \\
 664.26 & P(4)  & 4.304E-20 & 4.290E-20 & 0.32  \\
 676.01 & R(10) & 1.524E-19 & 1.519E-19 & 0.35  \\
 682.36 & R(18) & 1.679E-19 & 1.672E-19 & 0.43  \\
 690.36 & R(28) & 1.067E-19 & 1.061E-19 & 0.56  \\
 700.05 & R(40) & 3.230E-20 & 3.207E-20 & 0.70  \\
\\
2317.11 & P(36) & 1.000E-18 & 1.0031E-18 & -0.27   \\
2319.19 & P(34) & 1.238E-18 & 1.2402E-18 & -0.21   \\
2334.15 & P(18) & 3.278E-18 & 3.2818E-18 & -0.12   \\
2335.93 & P(16) & 3.329E-18 & 3.3318E-18 & -0.09   \\
2337.66 & P(14) & 3.277E-18 & 3.2795E-18 & -0.08   \\
2339.32 & P(12) & 3.112E-18 & 3.1145E-18 & -0.07   \\
2341.03 & P(10) & 2.831E-18 & 2.8323E-18 & -0.06   \\
2342.75 & P(8)  & 2.435E-18 & 2.4354E-18 & -0.02   \\
2344.38 & P(6)  & 1.933E-18 & 1.9337E-18 & -0.05   \\
2345.95 & P(4)  & 1.344E-18 & 1.3442E-18 & -0.01   \\
2362.79 & R(18) & 3.470E-18 & 3.4674E-18 &  0.07   \\
2369.08 & R(28) & 2.172E-18 & 2.1701E-18 &  0.09   \\
2371.42 & R(32) & 1.557E-18 & 1.5546E-18 &  0.13   \\
2372.55 & R(34) & 1.281E-18 & 1.2789E-18 &  0.16   \\
2373.66 & R(36) & 1.035E-18 & 1.0328E-18 &  0.19   \\
2374.74 & R(38) & 8.205E-19 & 8.1920E-19 &  0.16   \\
\hline
\end{tabular}
\end{table}

\bibliographystyle{naturemag}